\documentclass[12pt,epsf]{article}

\begin{document}

\begin{titlepage}

\begin{flushright}
\end{flushright}
\vskip 2.5cm

\begin{center}
{\Large \bf Perturbative Energy Shifts of the Luttinger Liquid Due to the
Presence of Additional Dimensions}
\end{center}

\vspace{1ex}

\begin{center}
{\large B. Altschul\footnote{current address: Department of Physics, Indiana
University, Bloomington, IN 47405; {\tt baltschu@mit.edu}}}

\vspace{5mm}
{\sl Department of Mathematics} \\
{\sl Massachusetts Institute of Technology} \\
{\sl Cambridge, MA, 02139-4307 USA} \\

\end{center}

\vspace{2.5ex}

\medskip

\centerline {\bf Abstract}

\bigskip

We consider a system of spinless fermions in a nearly one-dimensional
cylindrical trap. We introduce four-fermion interaction terms through which
particles in low-lying modes of the trap may interact with particles in
radially or angularly excited states. Treating these interactions as
perturbations about a purely one-dimensional Luttinger liquid, we calculate the
zero-temperature
energy shifts of the bosonic quanta of the Luttinger model. We perform this
calculation is two ways. First we use the bosonized forms of the fermion
operators. Then we introduce a simplified,
physically-motivated model; in this model, the methods used immediately
generalize, so that we may also
calculate the energy shifts of the Tomonaga-Luttinger model. By equating
the results from the two disparate models, we obtain the numerical
value of the Luttinger liquid bandwidth for this system.

\bigskip 

\end{titlepage}

\newpage

\section{Introduction: fermions trapped in one dimension}

A low-temperature system of fermions in one dimension has the known property
of bosonization~\cite{ref-schulz,ref-solyom,ref-emery1}. The coherent
fermion-hole excitations of the system behave
like bosons, with a spectrum that remains exactly solvable even in the presence
of certain arbitrarily strong interactions. This phenomenon is described by
the Luttinger and Tomonaga-Luttinger models.

The main physical applications of the Luttinger model occur in the context of
the quantum Hall effect~\cite{ref-wen1,ref-wen2}. However, there is another
situation in which we expect
Luttinger-type behavior to arise quite naturally. If a large number of
degenerate fermions are trapped in a
potential well with a very high aspect ratio, so that the particles are
effectively constrained to move along a single axis, we would expect
one-dimensional behavior to dominate. We shall consider a system of this type.
Since the trapping potential
cannot be purely one-dimensional, the second and third dimensions will
affect the energy levels of the system. We shall calculate these energy
shifts, by
treating the effects of the small transverse dimensions as perturbations around
a purely one-dimensional model. Since the field of fermion trapping is
advancing very
rapidly~\cite{ref-demarco,ref-schreck,ref-granade,ref-hadz},
such calculations should
be highly relevant in the analysis of future research.

After a brief discussion of trapped fermions and the one-dimensional Luttinger
model, we shall
introduce a physically motivated interaction Hamiltonian in
Section~\ref{sec-interaction}. We then calculate the perturbative energy
shift due to this interaction in two different ways. In
Section~\ref{sec-bosonized}, we use bosonized forms for the fermion
operators. The results of the bosonized calculation depend upon the cutoff of
the model. We may determine the physically meaningful value of the cutoff by
comparing the bosonized result with the result of our second calculation
(Section~\ref{sec-physical}), which uses an entirely different, simplified
model of the one-dimensional Fermi gas. Finally, in Section~\ref{sec-ext}, we
extend our results to systems with additional interactions and discuss the
significance of our findings. Our main result is expression
(\ref{eq-realshift}), which describes the energy shifts
of the bosonic modes.

We shall consider a system with a large number of fermions trapped
in a cylindrical potential. In the cylidrical coordinates, $(z,\theta,\rho)$,
the wave function is periodic in $z$ with period $L$ and has a Dirichlet
boundary condition at $\rho=b\ll L$. The boundary condition for $\rho$
corresponds to a trap with hard walls, but our general treatment will also
be applicable to other forms for the trapping potential. The eigenfuctions
of this trap are (setting $\hbar=1$)
\begin{equation}
u_{k,\ell,n_{\rho}}=\frac{e^{ikz}}{\sqrt{L}}\frac{e^{i\ell\theta}}{\sqrt{2\pi}}
\frac{J_{\ell}\left(j_{\ell,n_{\rho}}\rho/b\right)}{b\sqrt{-J_{\ell+1}(j_{\ell,
n_{\rho}})J_{\ell-1}(j_{\ell,n_{\rho}})/2}},
\end{equation}
where $k=\frac{2\pi r}{L}$ for some integer $r$, $\ell$ is an integer, and
$n_{\rho}$ is a positive integer, which represents the number of radial
nodes in the wave function. The number $j_{\ell,n_{\rho}}$ is
the $n_{\rho}$-th zero of the Bessel function $J_{\ell}$. The energies of
these modes are
\begin{equation}
E_{k,\ell,n_{\rho}}=\frac{1}{2m}\left[k^{2}+\left(\frac{j_{\ell,n_{\rho}}}{b}
\right)^{2}\right],
\end{equation}
so there are energy  gaps $\Delta_{\ell,n_{\rho}}=\frac{(j_{\ell,n_{\rho}})
^{2}-(j_{0,0})^{2}}{2mb^{2}}$ betwen the $n_{\rho}=\ell=0$
states and the higher exicited states.

In this paper, we are interested in the situation where the Fermi level is low
enough so that only the $n_{\rho}=\ell=0$ states are occupied, and the exicted
states of the trap can enter only virtually. This requires that the transverse
size $b$ of the trap be of the order of the mean spacing between adjacent
particles or smaller. (and most of our calculations will also require that the
temperature
of the system be vanishing). When no radially or angularly excited states are
occupied, $k$ becomes the only relevant quantum number, and the system becomes
effectively one-dimensional. This reduction
in dimensionality allows us to make use of the peculiar theory of
one-dimensional degenerate fermions, with its property of bosonization.

The phenomenon of bosonization is expressed in the Luttinger model, in which
the noninteracting (spinless), one-dimensional fermionic Hamiltonian is
approximated by
\begin{equation}
H_{0}=v_{F}\sum_{k}\left[(k-k_{F})a^{\dag}_{k}a_{k}+(-k-k_{F})b^{\dag}_{k}b_{k}
\right],
\end{equation}
where $a$ ($b$) operators correspond to right- (left-)moving fermions, and
$v_{F}$ is the Fermi velocity. The linearization of the
energy spectrum comes from the fact that the energy required to excite a
particle near the Fermi surface is $\frac{(k_{F}+q)^{2}}{2m}-\frac{k_{F}^{2}}
{2m}=v_{F}q+\frac{q^{2}}{2m}$; if the excitation momentum $q$ is small compared
to the Fermi momentum, them we may neglect the $q^{2}$ term.
In the
Luttinger model, the further approximation is made that this Hamiltonian is
valid for all momenta $k$, but that states corresponding to $a_{k}$ and
$b_{-k}$ for $k<0$ are filled~\cite{ref-luttinger,ref-mattis1}.  The
introduction of fictional
negative-energy states, which lie far from the Fermi
surface, should have little effect on the physics at low temperatures.

With these approximations, the Hamiltonian may be expressed in the form
\begin{equation}
H_{0}=\frac{2\pi v_{F}}{L}\sum_{q>0}[\rho_{+}(q)\rho_{+}(-q)+\rho_{-}(-q)
\rho_{-}(q)],
\end{equation}
where the $\rho_{\pm}$ operators
\begin{eqnarray}
\label{eq-bosonization}
\rho_{+}(q) & = & \sum_{k}a^{\dag}_{k+q}a_{k} \\
\rho_{-}(q) & = & \sum_{k}b^{\dag}_{k+q}b_{k}
\end{eqnarray}
obey the Bose commutation relations
$[\rho_{+}(-q),\rho_{+}(q')]=[\rho_{-}(q),\rho_{-}(-q')]=\delta_{qq'}
\frac{qL}{2\pi}$ and $[\rho_{+}(q),\rho_{-}(q')]=0$. The bosonic quanta
created and annihilated by the $\rho_{\pm}$ operators
have energies $\omega_{q}=v_{F}|q|$. It is also possible to
express the fermion field operators in terms of the bosonic $\rho_{\pm}$
operators. We shall make use of this representation in
Section~\ref{sec-bosonized}.

The Tomonaga-Luttinger model is formed by adding scattering 
effects~\cite{ref-tomonaga}, through
the interaction Hamiltonian
\begin{equation}
H_{1}=\frac{1}{2L}\sum_{q}\{2g_{2}(q)\rho_{+}(q)\rho_{-}(-q)+g_{4}(q)
[\rho_{+}(q)\rho_{+}(-q)+\rho_{-}(q)\rho_{-}(-q)]\}.
\end{equation}
The $g_{2}$ and $g_{4}$ are the Fourier transforms of the interaction
potentials. This interaction (like the similar interactions to be
introduced in Section~\ref{sec-interaction}) describes forward scattering
processes between fermions, with momentum exchange $q$.
The Tomonaga-Luttinger model is exactly solvable, since
$H_{0}+H_{1}$ remains bilinear
in the $\rho_{\pm}$. We shall return to this interaction in
Section~\ref{sec-ext}, but first we shall consider interactions between the
Luttinger liquid bosons and fermions in excited states of the trap.

\section{Interaction Hamiltonians}

\label{sec-interaction}

The Luttinger model should be a good description of a
system of fermions in a narrow, axially symmetric trap, in which no radially or
angularly excited states are occupied.
We shall add to this system additional two-fermion interaction terms. These
new interactions will be similar in structure
to $H_{1}$ (when $H_{1}$ is expanded in terms of fermion operators).
These interactions must conserve linear momentum along and angular momentum
around the
axis of the trap, as well as fermion number. One such interaction term is
\begin{equation}
H'=\sum_{k,k',q}C_{k,k',q}c^{\dag}_{k-q}d_{k}(a^{\dag}_{k'+q}a_{k'}+
b^{\dag}_{k'+q}b_{k'}).
\end{equation}
The operators $c_{k}$ and $d_{k}$ correspond to arbitrary fermion states,
subject to the condition that at least one of them corresponds to an excited
state of the trap (i.e. is not an $a_{k}$ or a $b_{k}$). This ensures that we
shall
be studying new effects. The effective coupling $C_{k,k',q}$ will depend upon
the radial and angular quantum numbers corresponding to $c_{k}$ and $d_{k}$
(in part because $C_{k,k',q}$ includes integrals over the radial and angular
eigenmodes of the trap). Angular momentum conservation demands that $c_{k}$
and $d_{k}$ annihilate states with the same angular momentum.

We are only interested in interactions that will renormalize the bosonic
energies. The simplest way for bosonized operators to arise from from an
interaction Hamiltonian such as $H'$ is for $C_{k,k',q}$ to be effectively
independent of $k$ and $k'$.  Then the $a^{\dag}_{k'+q}$ and $a_{k'}$ operators
combine to form a boson operator, as in (\ref{eq-bosonization}),
and the $b^{\dag}_{k'+q}$ and $b_{k'}$ combine similarly. This leads
to an interaction of the form
\begin{equation}
\label{eq-H2}
H_{2}=\frac{1}{2L}\sum_{q\neq0}g(q)\left(\sum_{k}c^{\dag}_{k-q}d_{k}\right)
\left[\rho_{+}(q)+\rho_{-}(q)\right]+{\rm h.c.}.
\end{equation}

We shall calculate the energy shifts of the low-lying bosonic states to
second order in $H_{2}$. (The first-order shifts obviously vanish.) If both
$c_{k-q}$ and $d_{k}$ correspond to excited states of the trap, then the
operator $H_{2}$ preserves the number of excited fermions. Hence, $H_{2}$ has
zero matrix element between low-lying (Luttinger liquid) states and states with
excited fermions. So we only get nonvanishing
contributions from terms where one of
$c_{k}$ and $d_{k}$ corresponds to an excited fermion state, while the other
corresponds to a low-lying mode. Moreover, the excited mode must have angular
momentum $\ell=0$ about the axis of the trap.

Let us fix $q>0$. We shall calculate the energy shift for the $\rho_{+}(q)$
quanta. In order for these quanta to be well-behaved Luttinger liquid phonons,
we must have $\omega_{q}\ll\epsilon_{F}\equiv\frac{1}{2m}k_{F}^{2}$, or
$q\ll k_{F}$. We may write the interactions affecting the $\rho_{+}(q)$ quanta
as
\begin{equation}
\label{eq-interaction}
H_{2,+q}=\frac{1}{L}g(q)\left\{\left[\sum_{k}(c^{\dag}_{k-q}d_{k}+d^{\dag}
_{k-q}c_{k})\right]\rho_{+}(q)+\left[\sum_{k}(c^{\dag}_{k+q}d_{k}+d^{\dag}
_{k+q}c_{k})\right]\rho_{+}(-q)\right\},
\end{equation}
where $d_{k}$ annihilates an excited fermion and $c_{k}$ annihilates a
low-lying fermion. [If $c_{k}$ annihilates a right-moving fermion, additional
terms are actually be required in order make $H_{2,+q}$ hermitian. However,
these terms do not involve $\rho_{+}$, so their matrix elements are smaller by
a factor of ${\cal O}(L^{-1/2})$ and may be neglected.] We shall denote by
$\Delta$ the energy gap $\Delta_{0,n_{\rho}}$ for the particular $\ell=0$
states corresponding to the operators $d_{k}$.

\section{Energy shift calculated with bosonized operators}

\label{sec-bosonized}

We shall calculate the perturbative energy shift arising from
(\ref{eq-interaction}) in two different models. In this section, we use
the bosonized
field operators of the Luttinger model to perform the calculation.
It is possible to obtain a closed-form expression for the energy shift in
two different situations---when the temperature vanishes or when the energy
gap is very large. (In each of these cases, the ``transverse temperature,''
which is the effective temperature for excitations of the trap's radial and
angular modes, is zero.) In these two situations, we
find that the leading contributions to the the infinite sum over intermediate
states may be resummed. This resummation allows us to find a simple expression
for the energy shift.

We shall restrict our attention to the case in which $c_{k}$ is $a_{k}$ (rather
than $b_{k}$). In this case, the matrix element which we must calculate is
$\langle\{m_{+p}\};K|H_{2,+q}|\{n_{+p}\}\rangle$. The state $|\{n_{+p}\}
\rangle$ is a number state of the Luttinger model; the $\{n_{+p}\}$ are the
occupation numbers of the $\rho_{+}$ quanta. The state $|\{m_{+p}\};K\rangle$
has differing occupation numbers $\{m_{+p}\}$, as well as an excited state
fermion with momentum $K$.

The matrix element is simply
\begin{eqnarray}
\langle\{m_{+p}\};K|H_{2,q}|\{n_{+p}\}\rangle & = & \frac{1}{L}g(q)\langle\{m
_{+p}\};K|\sum_{k}\left[d^{\dag}_{k-q}a_{k}\rho_{+}(q)
+d^{\dag}_{k+q}a_{k}\rho_{+}(-q)\right]|\{n_{+p}\}\rangle
\nonumber\\
& = & \frac{1}{L}g(q)\sum_{k}\delta_{K,k-q}\sqrt{\left(\frac{qL}{2\pi}\right)
\left(n_{+q}+1\right)}\langle\{m_{+p}\}|a_{k}|\{n'_{+p}\}\rangle \nonumber\\
\label{eq-Kmelement}
& & +\frac{1}{L}g(q)\sum_{k}\delta_{K,k+q}\sqrt{\left(\frac{qL}{2\pi}\right)n_
{+q}}\langle\{m_{+p}\}|a_{k}|\{n''_{+p}\}\rangle,
\end{eqnarray}
where the occupation numbers $\{n'_{+p}\}$ and $\{n''_{+p}\}$ are the same
as the $\{n_{+p}\}$, except that $n'_{+q}=n_{+q}+1$ and  $n''_{+q}=n_{+q}-1$.
As $L\rightarrow\infty$, $\{n'_{+p}\}$ and $\{n''_{+p}\}$ differ from 
$\{n_{+p}\}$ only on a set of measure zero, and we may replace the matrix
elements $\langle\{m_{+p}\}|a_{k}|\{n'_{+p}\}\rangle$ and
$\langle\{m_{+p}\}|a_{k}|\{n''_{+p}\}\rangle$ appearing in (\ref{eq-Kmelement})
by $\langle\{m'_{+p}\}|a_{k}|\{n_{+p}\}\rangle$ and
$\langle\{m''_{+p}\}|a_{k}|\{n_{+p}\}\rangle$, where $m'_{+q}=m_{+q}-1$ and
$m''_{+q}=m_{+q}+1$. We shall eventually be summing over all
possible sets of occupation numbers $\{m_{+p}\}$, so we may
ultimately replace both matrix elements in (\ref{eq-Kmelement})
with $\langle\{m_{+p}\}|a_{k}|\{n_{+p}\}\rangle$.
So we see that the crucial quantity to calculate is
$\langle\{m_{+p}\}|a_{k}|\{n_{+p}\}\rangle$, for an arbitrary set of occupation
numbers $\{m_{+p}\}$. We shall perform this calculation by taking a Fourier
transform,
\begin{equation}
\label{eq-FT}
\langle\{m_{+p}\}|a_{k}|\{n_{+p}\}\rangle=\frac{1}{\sqrt{L}}\int_{0}^{L}dx\,
e^{-ikx}\langle\{m_{+p}\}|\psi_{+}(x)|\{n_{+p}\}\rangle,
\end{equation}
where $\psi_{+}$ is the field operator for the right-moving fermions (while
$\psi_{-}$ is the left-moving fermions' field operator).
So we must now turn our attention to the operators $\psi_{\pm}$.

The bosonized form of the fermion field operator
is~\cite{ref-mattis2,ref-luther1,ref-heidenreich,ref-haldane1,ref-haldane2}
\begin{equation}
\psi_{\pm}(x)=\frac{1}{\sqrt{2\pi\alpha}}U_{\pm}\exp
[\pm ik_{F}x\mp i\phi(x)+i\theta(x)].
\end{equation}
In the pure Luttinger model, this is only an operator identity when we take the
limit $\alpha\rightarrow0$~\cite{ref-theumann}.
However, we shall interpret $\alpha$ as a finite
cutoff, so that $v_{F}\alpha^{-1}$ is the bandwidth of the Luttinger
liquid~\cite{ref-solyom,ref-luther1,ref-luther2,ref-lee,ref-chui}.
The finiteness of $\alpha$ corresponds to a deviation of the energy spectrum
from
the Luttinger form far from the Fermi surface. We shall discuss which values of
$\alpha$ are physically meaningful in Sections~\ref{sec-physical}
and~\ref{sec-ext}. The $U_{\pm}$ operators
decrease the total particle number on their corresponding branches by unity;
these operators are necessary because the bosonic fields conserve the total
fermion number.

The bosonic quantities $\phi$ (the boson field) and $\theta$ (the
integral of the momentum conjugate to $\phi$) are given by
\begin{eqnarray}
-i\phi & = & -\frac{\pi}{L}\sum_{p\neq0}\frac{1}{p}e^{-\alpha|p|/2}e^{-ipx}
[\rho_{+}(p)+\rho_{-}(p)]+i\pi\frac{N_{+}+N_{-}}{L}x \\
i\theta & = & -\frac{\pi}{L}\sum_{p\neq0}\frac{1}{p}e^{-\alpha|p|/2}e^{-ipx}
[\rho_{+}(p)-\rho_{-}(p)]+i\pi\frac{N_{+}-N_{-}}{L}x,
\end{eqnarray}
where $N_{+}$ and $N_{-}$ are the excess numbers of right- and left-moving
fermions relative to the ground state, respectively.
So the argument of the exponential in $\psi_{\pm}$ is
\begin{equation}
\pm ik_{F}x\mp i\phi+i\theta=\pm ik_{F}x\mp\frac{2\pi}{L}\sum_{p\neq0}\frac{1}
{p}e^{-\alpha|p|/2}e^{-ipx}\rho_{\pm}(p)\pm \frac{2\pi iN_{\pm}}{L}x.
\end{equation}
For states sufficiently close to the ground state, we may neglect the $N_{\pm}$
terms; this will be equivalent to negelcting an ${\cal O}(L^{-1})$ term in the
argument of a momentum-conserving $\delta$-function. So we have
\begin{equation}
\label{eq-phase}
\pm ik_{F}x\mp i\phi+i\theta=\pm ik_{F}x\mp\frac{2\pi}{L}\sum_{p>0}\frac{1}{p}
e^{-\alpha p/2}\left[e^{-ipx}\rho_{\pm}(p)-e^{ipx}\rho_{\pm}(-p)\right].
\end{equation}

Since $[\rho_{\pm}(p),\rho_{\pm}(p')]=0$ if $p\neq\pm p'$, the elements of the
sum in (\ref{eq-phase}) all commute with one-another. This allows us to write
$\psi_{\pm}$ in the product form
\begin{equation}
\psi_{\pm}(x)=\frac{1}{\sqrt{2\pi\alpha}}U_{\pm}e^{\pm
ik_{F}x}\prod_{p>0}\exp{\left\{\pm\left(\frac{2\pi}{pL}\right)
e^{-\alpha p/2}\left[e^{ipx}\rho_{\pm}(-p)-e^{-ipx}\rho_{\pm}(p)\right]
\right\}}.
\end{equation}
So the study of the matrix elements of $\psi_{\pm}$ between different Luttinger
liquid states leads naturally to the study of harmonic oscillator matrix
elements of the form $\langle m|e^{\lambda A^{\dag}-\lambda^{*}A}|n\rangle$,
where $A^{\dag}$ and $A$ are raising and lowering operators.
In our case, since $\sqrt{2\pi/(pL)}\rho_{\pm}$ is a canonically
normalized ladder operator,
\begin{equation}
\lambda=\mp\sqrt{\frac{2\pi}{pL}}e^{\mp ipx}e^{-\alpha p/2}.
\end{equation}
It is important to note that $\lambda$ is ${\cal O}(L^{-1/2})$.

In terms of $l\equiv m-n$, the general formula for the matrix element in
question is
\begin{eqnarray}
\langle m|e^{\lambda A^{\dag}-\lambda^{*}A}|n\rangle & = & \frac{1}{|l|!}
\left(\frac
{m!}{n!}\right)^{\frac{1}{2}{\rm sgn}(l)}\lambda^{(|l|+l)/2}(-\lambda^{*})
^{(|l|-l)/2} \nonumber\\
\label{eq-HOelements}
& & \times\left\{e^{|\lambda|^{2}/2}F\left[\max(m,n)+1;|l|+1;-|\lambda|^2
\right]\right\},
\end{eqnarray}
where $F(a;b;z)$ is a confluent hypergeometric function. This formula is
derived in Appendix~\ref{sec-appHO}.

The formulas
above pertain to both $\psi_{+}$ and $\psi_{-}$, but we shall henceforth
restrict our attention to $\psi_{+}$ only. We shall now proceed to calculate
the energy shift for the case in which the temperature is effectively zero
when compared to the transverse energy scale of the trap.
The fact that the temperature
vanishes allows us to make some substantial simplifications of the
mathematics. We shall show explicitly that these simplifications are not valid
(and would give rise to unphysical results) when the ``transverse temperature''
is nonzero.

It may initially appear that the bracketed factors in (\ref{eq-HOelements})
may be neglected, since both $e^{|\lambda|^{2}/2}$ and $F[\max(m,n)+1;|l|+1;
-|\lambda|^2]$ are power series in $|\lambda|^{2}$, and $|\lambda|^{2}$ is 
${\cal O}(L^{-1})$. Each power series is dominated by its first term in the
large $L$ limit, and for both series, that term is unity. However, these
terms may not actually be neglected in the $L\rightarrow\infty$ limit, because
there are an infinite number of such terms, corresponding to all the allowed
momenta. Since the density of states in momentum space is proportional to $L$,
the infinite product does not approach unity.

Let us denote the infinite product in question by ${\cal P}$; that is,
${\cal P}$ is defined to be
\begin{equation}
{\cal P}=\prod_{p>0}\exp\left[\left(\frac{\pi}{pL}\right)e^{-\alpha p}\right]
F\left[\max\left(m_{+p},n_{+p}\right)+1;\left|m_{+p}-n_{+p}\right|+1;-\left(
\frac{2\pi}{pL}\right)e^{-\alpha p}\right].
\end{equation}
When we calculate the energy shift, we sum up terms consisting of a matrix
element squared divided by an energy difference. If the
characteristic energy differences are
${\cal O}(L)$, then the sum of the matrix elements squared must also
be ${\cal O}(L)$ if we are to obtain a nonvanishing contribution. This is only
possible if the occupation numbers of a macroscopic number of states
states are ${\cal O}(L)$. This is a natural situation at finite temperature;
however, at $T=0$, the occupation numbers must be small, so that the
excitation energy of the system is not extensive.

So at $T=0$ we only get a nonvanishing contribution when the difference in
energies between $|\{m_{+p}\}\rangle$ and $|\{n_{+p}\}\rangle$ grows more
slowly than $L$. That is, almost all of the $m_{+p}$ must be equal to the
$n_{+p}$. We may then separate out those values of $p$ at which $m_{+p}$ and
$n_{+p}$ differ and get
\begin{eqnarray}
{\cal P} & = & \prod_{p,l_{+p}\neq0}\left\{\exp\left[\left(\frac{\pi}
{pL}\right)e^{-\alpha p}\right]
F\left[\max\left(m_{+p},n_{+p}\right)+1;\left|l_{+p}\right|+1;-\left(
\frac{2\pi}{pL}\right)e^{-\alpha p}\right]\right\} \nonumber\\
\label{eq-psplit}
& & \times\prod_{p,l_{+p}=0}\left\{\exp\left[\left(\frac{\pi}{pL}
\right)e^{-\alpha p}\right]F\left[n_{+p}+1;1;-\left(\frac
{2\pi}{pL}\right)e^{-\alpha p}\right]\right\}.
\end{eqnarray}
The number of terms in the first product must grow more slowly than $L$. Since
each of these terms has the form $1+{\cal O}(L^{-1})$ as
$L\rightarrow\infty$, the entire product is one in that limit.

The second product in (\ref{eq-psplit}) is more complicated. Using the Kummer
transformation formula $F(a;b;z)=e^{z}F(b-a;b;-z)$ and the fact that
$F(-n;1;z)$ is just the Laguerre polynomial $L_{n}(z)$~\cite{ref-abram}, we may
rewrite
${\cal P}$ as
\begin{equation}
{\cal P}=\prod_{p,l_{+p}=0}\left\{\exp\left[-\left(\frac{\pi}{pL}\right)
e^{-\alpha p}\right]L_{n_{+p}}\left[\left(\frac{2\pi}{pL}\right)e^{-\alpha p}
\right]\right\}
\end{equation}
We may evaluate this explicitly at $T=0$, where $n_{+p}=0$ for almost all $p$.
The product of the Laguerre polynomials is one, and all that remain are the
exponentials. Each momentum $p$ represented in the product has the form
$p(r)=2\pi r/L$, for some positive integer $r$, so we have
\begin{equation}
{\cal P}(T=0)=\exp\left[-\sum_{r,l_{+p(r)}=0}\frac{e^{-2\pi\alpha r/L}}
{2r}\right].
\end{equation}
Since $l_{+p}=0$ for almost all $p$, we may extend the sum to include all
positive integers $r$. Then the sum is just the Taylor series for $\frac{1}{2}
\log\left(1-e^{-2\pi\alpha/L}\right)$, so, to leading order in $L^{-1}$, we
have
\begin{equation}
{\cal P}(T=0)=\sqrt{\frac{2\pi\alpha}{L}}.
\end{equation}
The factor of $\sqrt{\frac{\alpha}{L}}$ will be
necessary in order to make our final result finite.

If we were to assume that the only terms which contributed had $m_{+p}=n_{+p}$
for almost all $p$ even when $T>0$, we would obtain the expression
\begin{equation}
{\cal P}=\sqrt{\frac{2\pi\alpha}{L}}\left\{\prod_{p,l_{+p}=0}L_{n_{+p}}\left[
\left(\frac{2\pi}{pL}\right)e^{-\alpha p}\right]\right\}.
\end{equation}
We shall use this expression to demonstrate explicitly that our assumption is
inconsistent unless $T$ vanishes.

We may now turn our attention to remaining factors in the matrix
elements---those arising from the unbracketed terms in (\ref{eq-HOelements}).
We shall divide the matrix elements of
this operator into groups. Each matrix element 
$\langle\{m_{+p}\}|\psi_{+}|\{n_{+p}\}\rangle$
is labeled by an integer $j\geq0$,
where $j$ is the number of quanta by which $\{m_{+p}\}$ and $\{n_{+p}\}$
differ. That is,
\begin{equation}
j=\sum_{p>0}|m_{+p}-n_{+p}|.
\end{equation}
According to our earlier argument, at $T=0$, the relevant values of $j$ are
smaller than ${\cal O}(L)$, and we shall for the moment assume that this is
also the case at finite $T$ (although we shall eventually show that this
assumption is inconsistent).

Each of the $j$ quanta by which $\{m_{+p}\}$ and $\{n_{+p}\}$ differ
corresponds to a momentum $p_{i}>0$, $i\in1,2,\ldots,j$. Since $j$ is smaller
than ${\cal O}(L)$, in the continuum
limit all the $p_{i}$ are distinct, except on a set of measure zero. So we
shall only need to use (\ref{eq-HOelements}) in the case where $|l|\leq1$.
Keeping this in mind, we may write the matrix elements of $\psi_{+}$ as
\begin{equation}
\langle\{m_{+p}\}|\psi_{+}(x)|\{n_{+p}\}\rangle=
\frac{{\cal P}}{\sqrt{2\pi\alpha}}e^{ik_{F}x}\prod_{i=1}^{j}\left\{
\begin{array}{l}
-\sqrt{\frac{2\pi}{p_{i}L}}e^{-ip_{i}x}e^{-\alpha p_{i}/2}\sqrt{n_{+p_{i}}+1}
\\
\sqrt{\frac{2\pi}{p_{i}L}}e^{ip_{i}x}e^{-\alpha p_{i}/2}\sqrt{n_{+p_{i}}}
\end{array}
\right..
\end{equation}
The upper term applies if $m_{+p_{i}}>n_{+p_{i}}$ (i.e. if $m_{+p_{i}}=
n_{+p_{i}}+1$), and the lower term otherwise.
We note that this matrix element is ${\cal O}(L^{-j/2})$.
Such a matrix element can only give rise to a nonvanishing energy shift if it
is multiplied by an appropriate positive power of $L$.

We may find the corresponding matrix element of $a_{k}$ by taking the Fourier
transform (\ref{eq-FT}). We have
\begin{eqnarray}
\langle\{m_{+p}\}|a_{k}|\{n_{+p}\}\rangle & = &
\frac{{\cal P}}{\sqrt{2\pi\alpha}}\left(\frac{2\pi}{L}\right)^{j/2}\frac{1}
{\sqrt{L}}\int_{0}^{L}dx\,\exp\left(-ikx+ik_{F}x+\sum_{i=1}^{j}\mp ip_{i}x
\right) \nonumber\\
\label{eq-bosonizedak}
& & \times\exp\left(-\frac{1}{2}\alpha\sum_{i=1}
^{j}p_{i}\right)\prod_{i=1}^{j}\frac{1}{\sqrt{p_{i}}}\left\{
\begin{array}{l}
-\sqrt{n_{+p_{i}}+1} \\
\sqrt{n_{+p_{i}}}
\end{array}
\right..
\end{eqnarray}
For each $i$, we have two possibilities, depending upon the relative magnitudes
of $m_{+p_{i}}$ and $n_{+p_{i}}$. In each case, the upper (lower) sign in the
exponential corresponds to the upper (lower) square root.
The integral over $x$ just gives $L\delta_{k,k_{F}+\sum_{i=1}^{j}\mp
p_{i}}$. The matrix element squared is then
\begin{equation}
\left|\langle\{m_{+p}\}|a_{k}|\{n_{+p}\}\rangle\right|^{2} 
=\frac{{\cal P}^{2}}{2\pi\alpha}\left(\frac
{2\pi}{L}\right)^{j}L\delta_{k,k_{F}+\sum_{i=1}^{j}\mp p_{i}}
\label{eq-matsquared}
\exp\left(-\alpha
\sum_{i=1}^{j}p_{i}\right)\prod_{i=1}^{j}\frac{1}{p_{i}}\left\{
\begin{array}{l}
n_{+p_{i}}+1 \\
n_{+p_{i}}
\end{array}
\right..
\end{equation}

To find the perturbative energy shift due to terms of this form, we must
divide the matrix element squared
by the energy defect and integrate over all possible
momenta. The phase-space factors arising from the integrations will cancel out
the $j$ factors of $2\pi/L$ in (\ref{eq-matsquared}). The energy defect is
given by
\begin{equation}
\label{eq-bdefect}
E_{\{n'_{+p}\};k\mp q}-E_{\{n_{+p}\}}=
\Delta+\frac{1}{2m}\left[(k\mp q)^{2}-k_{F}^{2}\right]\pm\omega_{q}+\sum_{i=1}
^{j}\pm\omega_{p_{i}}.
\end{equation}
$\Delta+\frac{1}{2m}(k\mp q)^{2}$ is the energy of the excited fermion;
$\frac{1}{2m}k_{F}^{2}$ is the energy lost by the Luttinger liquid ground state
with the removal of one fermion; and $\pm\omega_{q}+\sum_{i=1}^{j}\pm\omega
_{p_{i}}$ is the energy change due to the phonons gained or lost in the
intermediate state. There are $2^{j}$ possible choices of signs in the sum over
the quanta created or annihilated by $a_{k}$.

The momentum range over which we must integrate extends from $e^{-\gamma}
p_{\min}\equiv e^{-\gamma}\frac{2\pi}{L}$ to $+\infty$. The
Euler-Mascheroni constant $\gamma$ enters when we change from a sum to an
integral, through the formula
$\lim_{m\rightarrow+\infty}\left(\sum_{r=1}^{m}\frac{1}{r}-\int_{1}^{m}\frac{dr}{r}\right)=\gamma$. We may then absorb this $\gamma$-dependence into the limits
of integration. When we integrate all the $p_{i}$ over this
range, we must also divide by $j!$. This factor arises from our overcounting of
intermediate states---permuting the $p_{i}$ does not give rise to a new state.

So the energy shift consists of terms of the form
\begin{eqnarray}
\label{eq-jterm}
\Delta E_{j} & = &-\frac{{\cal P}^{2}}{2\pi\alpha}
\frac{q}{2\pi L}|g(q)|^{2}\left(\frac{2\pi}{L}\right)
^{j}\sum_{k}\sum_{s=\pm1}\sum_{t=1}^{2^{j}}\frac{1}{j!}\left(\frac{L}{2\pi}
\right)^{j} \\
& \times &
\int_{\frac{p_{\min}}{e^{\gamma}}}^{+\infty}\left(\prod_{i=1}^{j}dp_{i}\,\frac
{e^{-\alpha p_{i}}}{p_{i}}\left\{
\begin{array}{c}
n_{+p_{i}}+1 \\
n_{+p_{i}}
\end{array}
\right\}\right)
\frac{L\delta_{k,k_{F}+\sum_{i=1}^{j}\mp p_{i}}\left(n_{+q}+\frac{s+1}{2}
\right)}{\Delta+\frac{1}{2m}\left[(k-sq)^{2}-k_{F}^{2}\right]+s\omega_{q}+
\sum_{i=1}^{j}\pm\omega_{p_{i}}} \nonumber
\end{eqnarray}
The sum over $t$ runs over the $2^{j}$ choices of sign (and of $n_{+p_{i}}+1$
or $n_{+p_{i}}$).
The integrand in (\ref{eq-jterm}) behaves as $p_{i}^{-1}$ for small values of
$p_{i}$ and as $p_{i}^{-3}$ for large values of $p_{i}$, so the integral is
dominated by the region where all the $p_{i}$ are small.

Near $p_{i}=p_{\min}$, $p_{i}$ and $\omega_{p_{i}}$ are each
${\cal O}(L^{-1})$,
so we may neglect $\sum_{i=1}^{j}\mp p_{i}$ and $\sum_{i=1}^{j}\pm\omega
_{p_{i}}$ unless $j$ is ${\cal O}(L)$ or greater. Since we have assumed that
$j$ is smaller than ${\cal O}(L)$,
we may drop $\sum_{i=1}^{j}\mp p_{i}$ and $\sum_{i=1}^{j}
\pm\omega_{p_{i}}$. We shall not drop the $\exp\left(-\alpha\sum_{i=1}^{j}p_{i}
\right)$ term; this factor is needed to provide a cutoff for the momentum
integral.

By dropping the two sums, we simplify the integral substantially. The
expression becomes
\begin{equation}
\Delta E_{j} = -\frac{{\cal P}^{2}}{4\pi^{2}
\alpha}\sum_{s=\pm1}\sum_{t=1}^{2^{j}}
\frac{|g(q)|^{2}q\left(n_{+q}+\frac{s+1}{2}\right)
\frac{1}{j!}
\prod_{i=1}^{j}\int_{p_{\min}/e^{\gamma}}^{+\infty}dp_{i}\,\frac
{e^{-\alpha p_{i}}}{p_{i}}\left\{
\begin{array}{c}
n_{+p_{i}}+1 \\
n_{+p_{i}}
\end{array}
\right\}}{\Delta+\frac{1}{2m}\left[
(k_{F}-sq)^{2}-k_{F}^{2}\right]+s\omega_{q}}.
\end{equation}
We may now perform the sum over $t$, to get
\begin{equation}
\Delta E_{j}=-
\frac{{\cal P}^{2}}{4\pi^{2}\alpha}\sum_{s=\pm1}
\frac{|g(q)|^{2}q\left(n_{+q}+\frac{s+1}{2}\right)
\frac{1}{j!}\left[\int_{p_{\min}/e^{\gamma}}^{+\infty}dp_{i}\,\frac
{e^{-\alpha p_{i}}}{p_{i}}\left(2n_{+p_{i}}+1\right)\right]^{j}}
{\Delta+\frac{1}{2m}\left[
(k_{F}-sq)^{2}-k_{F}^{2}\right]+s\omega_{q}}.
\end{equation}
Furthermore, we may easily sum up all the terms with different values of $j$,
getting the exponential function
\begin{equation}
\label{eq-Eplus}
\Delta E_{+}=-
\frac{{\cal P}^{2}}{4\pi^{2}\alpha}\sum_{s=\pm1}
\frac{|g(q)|^{2}q\left(n_{+q}+\frac{s+1}{2}\right)
\exp\left[\int_{p_{\min}/e^{\gamma}}^{+\infty}dp_{i}\,\frac{e^{-\alpha
p_{i}}}{p_{i}}\left(2n_{+p_{i}}+1\right)\right]}
{\Delta+\frac{1}{2m}\left[
(k_{F}-sq)^{2}-k_{F}^{2}\right]+s\omega_{q}}.
\end{equation}
(The subscript ``$+$'' in $\Delta E_{+}$ indicates that this energy shift only
includes the terms arising from $\psi_{+}$---i.e. the terms in which $c_{k}$ is
$a_{k}$.) It now remains to evaluate the integral and add the contributions
from the terms for which $c_{k}$ is $b_{k}$.

It is impossible to evaluate the integral for a general set of $n_{+p_{i}}$.
However, we may simplify our expression for the energy shift substantially by
separating the two terms in the integral and transforming the integral back
into an infinite sum. We have
\begin{equation}
\label{eq-intsplit}
\exp\left[\int_{\frac{p_{\min}}{e^{\gamma}}}^{+\infty}dp_{i}\,\frac{e^{-\alpha
p_{i}}}
{p_{i}}\left(2n_{+p_{i}}+1\right)\right]=\exp\left[2\sum_{p_{i}>0}\left(\frac
{2\pi}{p_{i}L}\right)e^{-\alpha p_{i}}n_{+p_{i}}
\right]\exp\left[\sum_{p_{i}>0}
\left(\frac{2\pi}{p_{i}L}\right)e^{-\alpha p_{i}}\right].
\end{equation}
The second factor on the right-hand side of (\ref{eq-intsplit}) contains
the same sum that we encountered when we evaluated ${\cal P}(T=0)$;
in fact, this term is exactly $[{\cal P}(T=0)]^{-2}$. So we define a new
infinite product ${\cal D}$ as
\begin{eqnarray}
{\cal D} & = &
{\cal P}^{2}\exp\left[\int_{\frac{p_{\min}}{e^{\gamma}}}^{+\infty}dp_{i}\,
\frac{e^{-\alpha p_{i}}}{p_{i}}\left(2n_{+p_{i}}+1\right)\right] \nonumber\\
& = &
\prod_{p_{i}>0}\left\{\exp\left[\left(\frac{2\pi}{p_{i}L}\right)e^{-\alpha
p_{i}}n_{+p_{i}}\right]L_{n_{+p_{i}}}\left[\left(\frac{2\pi}{p_{i}L}\right)e^
{-\alpha p_{i}}\right]\right\}^{2}.
\end{eqnarray}

${\cal D}$ is an overall factor multiplying our expression for $\Delta
E_{+}$. It is clear that ${\cal D}=1$ when $T=0$ (since almost all the $n_{+p}$
are then zero). For $T>0$, ${\cal D}$ contains all the energy shift's
temperature dependence, so it is natural to calculate its thermal average. We
shall find that this average diverges for all finite temperatures; this
unphysical
result signals that we can not assume that $j$ is smaller than ${\cal O}(L)$
unless the temperature is strictly vanishing.

Denoting the inverse temperature by $\beta$, we find that the thermal
average of ${\cal D}$ is
\begin{equation}
\langle{\cal D}\rangle={\cal N}\sum_{\{n_{+p}\}}\left\{\exp\left[-\beta v_{F}
\sum_{p>0}\left(pn_{+p}\right)\right]{\cal D}\right\}.
\end{equation}
The sum over the $\{n_{+p}\}$ runs over all possible sets of occupation
numbers, and ${\cal N}$ is the necessary normalization factor
\begin{equation}
{\cal N}=\left\{\sum_{\{n_{+p}\}}\exp\left[-\beta v_{F}
\sum_{p>0}\left(pn_{+p}\right)\right]\right\}^{-1}=\prod_{p>0}\left(1-e^{-\beta
v_{F}p}\right).
\end{equation}
$\langle{\cal D}\rangle$ may be factorized into an infinite product
of sums indexed by $p$. This allows us to calculate the thermal average
separately for each value of the momentum, so $\langle{\cal D}\rangle$ becomes
\begin{equation}
\langle{\cal D}\rangle={\cal N}\prod_{p>0}\sum_{n_{+p}=0}^{\infty}\left(\exp
\left[-\beta v_{F}pn_{+p}\right]\left\{\exp\left[\left(\frac{2\pi}{pL}
\right)e^{-\alpha p}n_{+p}\right]L_{n_{+p}}\left[\left(\frac{2\pi}
{pL}\right)e^{-\alpha p}\right]\right\}^{2}\right).
\end{equation}
The sum is now in the form of a product of generating functions for
$(L_{n})^{2}$.
That is,
\begin{equation}
\langle{\cal D}\rangle={\cal N}\prod_{p>0}\sum_{n=0}^{\infty}[L_{n}(x)]^{2}
z^{n},
\end{equation}
with $z=\exp\left[\left(\frac{2\pi}{pL}\right)e^{-\alpha p}-\beta v_{F}p
\right]$ and $x=\left(\frac{2\pi}{pL}\right)e^{-\alpha p}$.

The sum of this generating function is known~\cite{ref-gradsteyn}. In fact,
when $|z|<1$,
\begin{equation}
\sum_{n=0}^{\infty}[L_{n}(x)]^{2}z^{n}=\frac{1}{1-z}\exp\left(-\frac{2xz}{1-z}
\right)I_{0}\left(2x\frac{\sqrt{z}}{1-z}\right),
\end{equation}
where $I_{0}$ is the modified Bessel function $I_{0}(u)=J_{0}(iu)$. The series
diverges if $|z|>1$.
So $\langle{\cal D}\rangle$ converges only if $\exp\left[\left(\frac{2\pi}{pL}
\right)e^{-\alpha p}-\beta v_{F}p\right]<1$ for all $p$. Since $\left(\frac{2
\pi}{pL}\right)e^{-\alpha p}-\beta v_{F}p$ is a strictly decreasing function
of $p$, we need only consider $p=p_{\min}$. At this value of $p$, we may
neglect $e^{-\alpha p}$, so the condition for convergence becomes $\beta>\left(
v_{F}p_{\min}\right)^{-1}$. As $L\rightarrow\infty$, this can not be satisfied
at any finite temperature. So the thermal average of ${\cal D}$ is equal to
unity at $T=0$ (when $z=0$ for all $p$) and diverges for $T>0$. \{For finite
$L$ and very large $v_{F}$, the condition that $\beta>\left(v_{F}p_{\min}
\right)^{-1}$ gives us a qunatitative estimate of how small the termperature
must be in order for our approximation to be valid. Current experiments with
degenerate fermions~\cite{ref-demarco,ref-schreck,ref-granade,ref-hadz}
typically achieve
temperatures for which $\beta^{-1}\sim\epsilon_{F}$. In order for the $T=0$
approximation to be valid, the temperature must be smaller by a factor of
${\cal O}(N)$, so this regime is far beyond the limits of current technique.\}

The divergence of $\langle{\cal D}\rangle$ at nonzero $T$ indicates the
breakdown of our approximation. We may see this more explicitly by looking
at the argument of the exponential in (\ref{eq-Eplus}). The Taylor series
$\sum_{j=0}^{\infty}\frac{x^{j}}{j!}$ for $e^{x}$ is dominated by the terms
for which $j$ is ${\cal O}(x)$. At $T=0$, the argument of the exponential
appearing in (\ref{eq-Eplus}) is $\int_{\frac{p_{\min}}{e^{\gamma}}}^{+\infty}
dp_{i}\,\frac{e^{-\alpha p_{i}}}{p_{i}}=\log\left(\frac{L}{2\pi\alpha}\right)$,
which is ${\cal O}(\log L)$. Therefore, the dominant contributions to the
exponent come from values of $j$ that are likewise ${\cal O}(\log L)$.
However, when $T>0$, the average number of quanta present in each of the
lowest-lying modes is ${\cal O}(L)$ unless $\beta>(v_{F}p_{\min})^{-1}$. The
argument of the exponential and the dominant values of $j$ are then ${\cal O}
(L\log L)$. This contradicts our assumption that $j$ is smaller than ${\cal O}
(L)$.

We have shown that when the temperature is nonzero, we must receive
a substantial contribution from terms for which $j$ is ${\cal O}(L)$ or larger,
so the methods we have used so far do not apply to this case. However, the
$T=0$ treatment does provide us with several insights that will prove useful
when we analyze the case in which $T>0$ but the ``transverse temperature''
is still vanishing. We shall return to that case later; however,
there are several more points about the zero-temperature situation that we
must consider first.

Thus far, we have assumed that $\alpha^{-1}$ is the correct cutoff for the
momentum integral appearing in (\ref{eq-Eplus}). However, there is also another
cutoff we must consider. In order to obtain a physically meaningful result, we
must determine which cutoff is the relevant one.

A real Luttinger liquid has a finite bandwidth $v_{F}\alpha^{-1}$, because
the phonon description breaks down at large values of
$p_{i}$. The Luttinger model only describes the low-energy excitations of the
fermion system. When $p_{i}$ becomes comparable to $k_{F}$, the energy
$\frac{(k_{F}+p_{i})^{2}}{2m}-\frac{k_{F}^{2}}{2m}$ of an excitation can no
longer be approximated by $v_{F}p_{i}$ and is no longer much smaller than
$\epsilon_{F}$, so we expect the physical cutoff $\alpha^{-1}$ to be
${\cal O}(k_{F})$.

The second cutoff arises from the energy denominator (\ref{eq-bdefect}).
When $p_{i}$ is large compared
to $\sqrt{m\Delta}$, the integrand decays as $p_{i}^{-3}$. If there existed
an intermediate range of $p_{i}$ values, where $m\Delta/k_{F}\ll p_{i}\ll\sqrt
{m\Delta}$, the integrand would decay as $p_{i}^{-2}$ in this region. However,
since we must have $\Delta>\frac{1}{2m}k_{F}^{2}$ in order to have an energy
gap at all, such a region can not exist. So the rapid $p_{i}^{-3}$
decay at large values of $p_{i}$ will provide a cutoff for the integral which
we expect to be ${\cal O}(\sqrt{m\Delta})$.

We must determine which of these two cutoff scales is smaller, since the
smaller one determines the behavior of the energy shift.
When $\Delta\gg\epsilon_{F}$, the first cutoff, near the Fermi momentum, is
smaller. However, when the energy gap is comparable to the Fermi energy,
$\sqrt{m\Delta}$ is itself ${\cal O}(k_{F})$. Without more precise values for
the cutoffs, it is not clear which one matters in this situation.

We shall now estimate the cutoff $p'_{{\rm max}}$ arising from the
$p_{i}^{-3}$ behavior of (\ref{eq-bdefect}). To do this, we shall
again make use of the fact that the integral over the $p_{i}$ is dominated by
the region where the $p_{i}$ are small. The value of $p'_{{\rm max}}$ is
determined by the behavior of the integrand when not all the $p_{i}$ are small;
at least one of the $p_{i}$ must be comparable to $\sqrt{m\Delta}$. However,
the dominant contribution under such conditions comes from the region where
only a single $p_{i}$ is large compared to $p_{{\rm min}}$. If more than one
$p_{i}$ is large, the contribution is much smaller. This
means that we may determine $p'_{{\rm max}}$ by analyzing the case in which
all but one of the $p_{i}$ are negligible.

More precisely, we approximate $\delta_{k,k_{F}+\sum_{i=1}^{j}\mp p_{i}}$ by
$\delta_{k,k_{F}-p_{i}}$ and $\sum_{i=1}^{j}\pm\omega_{p_{i}}$ by
$\omega_{p_{i}}$. The choice of signs corresponds to the fact that the virtual
phonon at momentum $p_{i}$ must be created, not annihilated, since almost all
the phonon modes are empty of quanta when $T=0$. One of the $p_{i}$
integrations is now
nontrivial. We shall perform this integration explicitly and extract the cutoff
from it. Our prescription for doing this is that
\begin{equation}
\label{eq-cutofffind}
\int_{\frac{p_{\min}}{e^{\gamma}}}^{+\infty}\frac{dp_{i}}{p_{i}}\frac{1}{\Delta
+\frac{1}{2m}[(k_{F}-p_{i})^{2}-k_{F}^{2}]+\omega_{p_{i}}}\approx\frac{1}
{\Delta}\log\left(\frac{p'_{{\rm max}}}{p_{{\rm min}}}\right).
\end{equation}

The integral in (\ref{eq-cutofffind}) is elementary.
The result is
\begin{equation}
\int_{\frac{p_{\min}}{e^{\gamma}}}^{+\infty}\frac{dp_{i}}{p_{i}}\frac{1}
{\Delta+\frac{p_{i}^{2}}{2m}}=\frac{1}{\Delta}\log\left(\frac{e^{\gamma}}
{p_{\min}}\sqrt{2m\Delta+e^{2\gamma}p_{\min}^{2}}\right),
\end{equation}
so, neglecting $e^{2\gamma}p_{\min}^{2}$ compared to $2m\Delta$, we have that
$p'_{\max}=e^{\gamma}\sqrt{2m\Delta}$. (This is really the appropriate value of
$p'_{\max}$ only when $q\ll\sqrt{2m\Delta}$. A more careful calculation, which
takes into account the $q$-dependence of the cutoff, is located in
Appendix~\ref{sec-appcutoff}.)

Although this estimate is rather inexact, it shows the general character of the
cutoff. At $\Delta=\epsilon_{F}$, the cutoff takes on its minimum value of
$p'_{\max}\approx 1.78k_{F}$. However, we have not yet
determined whether this is ever the relevant cutoff.

We shall defer the final resolution of this question until
Section~\ref{sec-physical}. In the meantime, we shall make the replacement
\begin{equation}
\label{eq-prescription}
\int_{\frac{p_{\min}}{e^{\gamma}}}^{+\infty}dp_{i}\,\frac{e^{-\alpha p_{i}}}
{p_{i}}\rightarrow\log\left(\frac{Lp_{{\rm max}}}{2\pi}\right),
\end{equation}
where in all cases, $p_{{\rm max}}$ is ${\cal O}(k_{F})$.
Using the prescription (\ref{eq-prescription}) to evaluate the integral and
inserting our expression for ${\cal P}(T=0)$, we get
\begin{equation}
\label{eq-bosonizedE}
\Delta E_{+}\approx-\frac{1}{4\pi^{2}}\sum_{s=\pm1}
\frac{|g(q)|^{2}qp_{{\rm max}}\left(n_{+q}+\frac{s+1}{2}\right)}{\Delta+\frac
{1}{2m}\left[(k_{F}-sq)^{2}-k_{F}^{2}\right]+s\omega_{q}}.
\end{equation}
The denominator, $\Delta+\frac{1}{2m}[(k_{F}-sq)^{2}-k_{F}^{2}]+s\omega_{q}=
\Delta+\frac{q^{2}}{2m}$ is independent of $s$, so we may 
trivially perform the sum
over $s$. Thus far, we have restricted our attention to the case for which
$c_{k}$ corresponds to a right-moving fermion. The terms in which $c_{k}$
corresponds to a left-moving particle also contributes an equal amount,
so we have
\begin{equation}
\label{eq-bothbranchE}
\Delta E\approx-\frac{|g(q)|^{2}qp_{{\rm max}}}
{\pi^{2}\left(\Delta+\frac{q^{2}}{2m}\right)}\left(n_{+q}+\frac{1}{2}\right).
\end{equation}

(\ref{eq-bothbranchE}) is the energy shift for a single bosonic mode, with
momentum
$+q$. To obtain the total shift for the system, we must add in the similar
renormalization of the left-moving mode with momentum $-q$ and integrate over
$q$.  Since the density of momentum states is proportional to $L$, the
total shift is extensive, as it must be. We must also sum over all the
possible $n_{\rho}$ values to which $d_{k}$ may correspond; this will give
an infinite sum of terms of the form (\ref{eq-bothbranchE}), in which $\Delta$
takes on all the values $\Delta_{0,n_{\rho}}$, $n_{\rho}>0$.

We shall now turn our attention to the $T>0$ case. The crucial simplification
that allowed us to evaluate the $T=0$ energy shift in closed form was
the fact that the energy defects were effectively independent of the
intermediate states.
A remarkable cancellation then occurred between the prefactor ${\cal P}$
and the exponential that arose when we summed over all all intermediate states.
A similar cancellation will occur here.

When $T>0$, we may not necessarily neglect $\sum_{i=1}^{j}\mp p_{i}$ and
$\sum_{i=1}^{j}\pm\omega_{p_{i}}$, because $j$ is not generally small. The
presence of these two sums gives rise to a nontrivial energy denominator.
However, if $\Delta$ is formally very large, so that any other energy may
be neglected compared to it, the energy defect again becomes independent of the
$p_{i}$. As in the previous case, the temperature is negligible when compared
with the transverse energy scale; that is, the ``transverse temperature'' is
effectively vanishing.
With this simplification, it will again be possible to obtain an analytical
expression for the energy shift.

If we again assemble the various terms that compose the energy shift (now
assuming that $\Delta$ is very large), we have
\begin{equation}
\label{eq-largeDelta}
\Delta E_{+}=-\sum_{s=\pm 1}\frac{1}{2\pi\alpha}\frac{|g(q)|^{2}q\left(n_{+q}+
\frac{s+1}{2}\right)}{2\pi}\sum_{\{m_{+p}\}}\frac{\left|\langle\{
m_{+p}\}|\prod_{p>0}e^{\lambda_{+p}A_{+p}^{\dag}-\lambda_{+p}^{*}A_{+p}}|
\{n_{+p}\}\rangle\right|^{2}}{\Delta},
\end{equation}
where $\lambda_{+p}=-\sqrt{2\pi/(pL)}e^{-\alpha p/2}$, and $A_{+p}^{\dag}$ and
$A_{+p}$ are the canonical ladder operators corresponding to the $\rho_{+}(\pm
p)$ quanta.

Because the energy defect in (\ref{eq-largeDelta}) is independent of the
$\{m_{+p}\}$, we may rewrite the energy shift as
\begin{eqnarray}
\Delta E_{+} & = & -\sum_{s=\pm1}\frac{|g(q)|^{2}q\left(n_{+q}+
\frac{s+1}{2}\right)}{4\pi^{2}\alpha\Delta}
\langle \{n_{+p}\}|\left\{\prod_{p>0}\left(e^{\lambda_{+p}A_{+p}^{\dag}-\lambda
_{+p}^{*}A_{+p}}\right)^{\dag}\right. \nonumber\\
& & \times\left.\left[\sum
_{\{m_{+p}\}}|\{m_{+p}\}\rangle\langle\{m_{+p}\}|\right]\prod_{p>0}
\left(e^{\lambda_{+p}A_{+p}^{\dag}-\lambda_{+p}^{*}A_{+p}}\right)\right\}|
\{n_{+p}\}\rangle.
\end{eqnarray}
The operator in brackets is just the identity, so we may drop it. Furthermore,
since $\lambda_{+p}A_{+p}^{\dag}-\lambda_{+p}^{*}A_{+p}$ is antihermitian,
$\prod_{p>0}\left(e^{\lambda_{+p}A_{+p}^{\dag}-\lambda_{+p}^{*}A_{+p}}\right)
^{\dag}=\left[\prod_{p>0}\left(e^{\lambda_{+p}A_{+p}^{\dag}-\lambda_{+p}^{*}
A_{+p}}\right)\right]^{-1}$, and the entire matrix element expression is unity.
[We now see that the remarkable simplicity of the $T=0$ result is
a consequence of the fact that when the energy denominator is constant,
the sum over intermediate states may be evaluated simply using the closure
relation. However, the detailed analysis that led up to
(\ref{eq-bosonizedE}) was necessary for us to justify taking
the energy denominator to be constant.]

This $T>0$ result shows that, when $\Delta$ is very large, we recover a
similar expression to the one we obtained in a different extreme limit, that
of zero temperature. The total energy shift is
\begin{equation}
\label{eq-Tneq0E}
\Delta E=-\frac{|g(q)|^{2}q}{\pi^{2}\alpha\Delta}\left(n_{+q}+\frac{1}{2}
\right).
\end{equation}
This expression is actually identical to the large $\Delta$ limit of
(\ref{eq-bothbranchE}). In that limit, $\frac{q^{2}}{2m}$ may be neglected
compared to $\Delta$, and the correct cutoff $p_{\max}$ is just $\alpha^{-1}$,
because $p'_{\max}$ grows as ${\cal O}(\sqrt{m\Delta})$. However, we still do
not know to what actual value $\alpha$ corresponds. We shall now turn our
attention to a different model, which will allow us to answer that question.

\section{Energy shift for a more physical model}

\label{sec-physical}

In this section, we present an alternative model---one which is useful when
$T=0$ and $\Delta$ is very large. Using this model, we shall again calculate
the energy shift. From our calculation we may extract the physically relevant
value of $\alpha^{-1}$, the
cutoff arising from the breakdown of the phonon description.
Moreover, we shall find in Section~\ref{sec-ext} that our calculation requires
only slight modification in the presence of strong interactions between the
low-lying fermion states.

The Luttinger model does not accurately represent the fermions lying deep
within the Fermi sea, so it is not necessary to use a bosonized form
for the matrix elements of $c_{k}$ when $k\not\approx
k_{F}$. Since a basic assumption of the Luttinger liquid theory is that the
presence or absence of additional fermions far from the Fermi surface should
have minimal effect on the low-energy excitations, we shall simply assume that
the action of $c_{k}$ ($k\not\approx k_{F}$) on a Luttinger liquid state is to
create a hole in the Fermi sea, without affecting the phonon configuration in
any way. The only exception to this prescription is that when the fermion
operators are combined in the form (\ref{eq-bosonization}), as they are in
$H_{2}$, we shall used the bosonized expressions.

Our approximation will require
that $\Delta\gg\epsilon_{F}$. This condition ensures that the energy shift is
not dominated by the terms with the smallest energy defects; instead, the
defects
are dominated by $\Delta$ and depend only weakly on the energy of the fermion
annihilated by $c_{k}$. So the entire Fermi sea will contribute to the energy
shift, and the region near the Fermi surface (where our approximation for
$c_{k}$ is invalid) will make a comparatively small contribution. The condition
that $T=0$ ensures that the occupied states are precisely those with
$|k|<k_{F}$.

Now we must identify the states between which we shall be calculating matrix
elements. As before, we shall have the state $|\{n_{+p}\}\rangle$ on the
right. We shall find the energy shift of this state due to interactions with
the states $|\{n'_{+p}\};k\mp q;k\rangle$. These states have bosonic
occupation numbers $\{n_{+p}'\}$ which are the same as $\{n_{+p}\}$, except
that $n_{+q}'=n_{+q}\pm1$. These states also have an excited fermion with
momentum $k\mp q$ and a hole in the Fermi sea with momentum $-k$.

The matrix elements we shall need are then
\begin{equation}
\langle\{n'_{+p}\};k-sq;k|\rho_{+}(sq)d^{\dag}_{k-sq}c_{k}|\{n_{+p}\}\rangle=
\sqrt{\frac{qL}{2\pi}\left(n_{+q}+\frac{s+1}{2}\right)}\theta(k_{F}-|k|),
\end{equation}
for $s\pm 1$. The step function ensures that the matrix elements vanish when
$|k|>k_{F}$.
The energy defects are
\begin{equation}
E_{\{n'_{+p}\};k-sq;k}-E_{\{n_{+p}\}}=\Delta+\frac{1}{2m}\left[(k-sq)^{2}-k^{2}
\right]+s\omega_{q},
\end{equation}
so the energy shift is
\begin{eqnarray}
\label{eq-Eshift}
\Delta E & = & -\frac{1}{L^{2}}|g(q)|^{2}\sum_{k=-k_{F}}^{k_{F}}\sum_{s=\pm1}
\frac{|\langle\{n'_{+p}\};k-sq;k|\rho_{+}(sq)d^{\dag}_{k-sq}c_{k}|\{n_{+p}\}
\rangle|^{2}}{\Delta+\frac{1}{2m}\left[(k-sq)^{2}-k^{2}\right]+s\omega_{q}}
\nonumber\\
& = & -\frac{1}{L^{2}}|g(q)|^{2}\left(\frac{qL}{2\pi}\right)\sum_{k=-k_{F}}^{k
_{F}}\sum_{s=\pm1}\frac{n_{+q}+\frac{s+1}{2}}{\Delta+\frac{1}{2m}\left[(k-sq)
^{2}-k^{2}\right]+s\omega_{q}}.
\end{eqnarray}
Converting the sum over $k$ into an integral, we have
\begin{eqnarray}
\Delta E & = & -\frac{1}{L^{2}}|g(q)|^{2}\left(\frac{qL}{2\pi}\right)\sum_{s=
\pm1}\left(n_{+q}+\frac{s+1}{2}\right)\left(\frac{L}{2\pi}\right)\int_{-k_{F}}
^{k_{F}}dk\,\frac{1}{\left(\Delta+\frac{q^{2}}{2m}+s\omega_{q}\right)-s\frac{q}
{m}k} \nonumber\\
& = & -\frac{|g(q)|^{2}q}{4\pi^{2}}\sum_{s=\pm1}
\left(n_{+q}+\frac{s+1}{2}\right)\left(\frac{sm}{q}\right)
\log\left[\frac{\left(\Delta+\frac{q^{2}}{2m}+s\omega_{q}\right)+s\frac{q}{m}k
_{F}}{\left(\Delta+\frac{q^{2}}{2m}+s\omega_{q}\right)-s\frac{q}{m}k_{F}}
\right] \nonumber\\
\label{eq-logshift}
& = & -\frac{|g(q)|^{2}m}{4\pi^{2}}\sum_{s=\pm1}\left(n_{+q}+\frac{s+1}{2}
\right)s\log\left(1+\frac{2sqk_{F}/m}{\Delta+\frac{q^2}{2m}}\right).
\end{eqnarray}

We have already required that $\Delta$ be large compared with the Fermi energy.
Since $q\ll k_{F}$, we may use the approximation $\log(1+x)\approx x$,
approximating the logarithm in (\ref{eq-logshift}) by its leading term.
(In fact, because of cancellation between the $s=\pm1$ terms, this is
equivalent to approximating the logarithm by its two leading terms for
purposes of determining the frequency.) This gives us
\begin{equation}
\label{eq-realshift}
\Delta E=-\frac{|g(q)|^{2}qk_{F}}{\pi^{2}\left(\Delta+\frac{q^{2}}{2m}\right)}
\left(n_{+q}+\frac{1}{2}\right).
\end{equation}
As in Section~\ref{sec-bosonized}, we must sum over all values of $q$ to get an
extensive energy shift.

We now compare the result (\ref{eq-realshift}) with (\ref{eq-bothbranchE}).
The models leading to these expressions are both valid when $T=0$ and the
energy gap is large, so the two expressions must agree under these
circumstances. In fact, (\ref{eq-bothbranchE}) and (\ref{eq-realshift}) agree
in their region of common validity exactly if $p_{{\rm max}}=k_{F}$. Since
$k_{F}$ is substantially smaller than any possible value of $p'_{{\rm max}}$
(for large $\Delta$),
we must conclude that the relevant cutoff is $p_{\max}=\alpha^{-1}$.
The value of $\alpha^{-1}$ is a property solely of the Luttinger liquid, and
$\alpha^{-1}$ is thus idependent of $\Delta$. Therefore, $\alpha^{-1}$ is
always equal the Fermi momentum when $T=0$. We then see
that $\alpha^{-1}< p'_{{\rm max}}$ for all values of $\Delta$, so
$\alpha^{-1}$ is always the most relevant cutoff, and (\ref{eq-realshift}) is
the general expression for the $T=0$ energy shift.
(We shall discuss the possible relationship between $\alpha$ and the
temperature in
Section~\ref{sec-ext}.)

A simple estimate of the magnitude of any
correction to (\ref{eq-realshift}) due to $p'_{\max}$ is that the
fractional error should be approximately ${\cal O}(k_{F}/p'_{\max})$.
Since we found a minimum value of $p'_{\max}$ of only $1.78k_{F}$, this
error could be significant when the energy gap is small. This must be kept in
mind when this formula is applied to systems for which $\Delta$ is ${\cal O}
(\epsilon_{F})$.

\section{Extensions and discussion}

\label{sec-ext}

The model outlined in Section~\ref{sec-physical} may be applied to systems
with more general interactions. In particular,
our techniques may be applied to the interacting Tomonaga-Luttinger model.
The Tomo\-naga-Luttinger Hamiltonian $H_{0}+H_{1}$ is diagonalized by the
Bogoliubov-transformed operators~\cite{ref-bogoliubov}
\begin{equation}
\label{eq-tomonaga-rho}
\rho'_{\pm}(q)=\frac{1}{2}\left(X+X^{-1}\right)\rho_{\pm}(q)+\frac{1}{2}\left(
X-X^{-1}\right)\rho_{\mp}(q),
\end{equation}
where
\begin{equation}
X=\left(\frac{v_{F}+\frac{g_{4}(q)}{2\pi}+
\frac{g_{2}(q)}{2\pi}}{v_{F}+\frac{g_{4}(q)}{2\pi}-\frac{g_{2}(q)}{2\pi}}
\right)^{1/4}.
\end{equation}
The $\rho'_{\pm}$ satisfy the same commutation relations as the $\rho_{\pm}$.
To determine the effects of the additional interaction $H_{2}$, we only need
relate $\rho_{+}(q)+\rho_{-}(q)$ to $\rho'_{+}(q)+\rho'_{-}(q)$. From
(\ref{eq-tomonaga-rho}), it is obvious that
\begin{equation}
\label{eq-tomonaga-factor}
\rho_{+}(q)+\rho_{-}(q)=\left(\frac{v_{F}+\frac{g_{4}(q)}{2\pi}-\frac{g_{2}(q)}
{2\pi}}{v_{F}+\frac{g_{4}(q)}{2\pi}+\frac{g_{2}(q)}{2\pi}}\right)^{1/4}
[\rho'_{+}(q)+\rho'_{-}(q)],
\end{equation}
so the calculation of
$\Delta E$ requires only two modifications. We must use the
Tomonaga-Luttinger frequency
\begin{equation}
\omega'_{q}=|q|\left[\left(v_{F}+\frac{g_{4}}
{2\pi}\right)^{2}-\left(\frac{g_{2}}{2\pi}\right)^{2}\right]^{1/2}
\end{equation}
in place
of $\omega_{q}$,
and we must multiply the whole expression for the energy shift by $|X|^{-2}$.
This complicates the expressions somewhat; however, if we neglect terms
of ${\cal O}(\omega_{q})$ compared to $\Delta$ (since $q\ll k_{F}$), we simply
have
\begin{equation}
\label{eq-tomonagashift}
\Delta E=-\frac{|g(q)|^{2}qk_{F}}{\pi^{2}\Delta}
\left|\frac{v_{F}+\frac{g_{4}(q)}{2\pi}-\frac{g_{2}(q)}
{2\pi}}{v_{F}+\frac{g_{4}(q)}{2\pi}+\frac{g_{2}(q)}{2\pi}}\right|^{1/2}
\left(n_{+q}+\frac{1}{2}\right).
\end{equation}
It is slightly more difficult to perform this calculation using the
bosonized operators. The mixing of the the
right-moving and left-moving phonons in $\psi_{\pm}$ adds to the complexity of
the calculations~\cite{ref-solyom,ref-haldane2,ref-fogedby,ref-emery2}.
Performing the calculation, we find that the matrix elements factorize into
$\rho'_{+}$ terms and $\rho'_{-}$ terms. The cancellations that appeared in
Section~\ref{sec-bosonized} at $T=0$
now occur separately for the two types of terms.
The end result is the same as (\ref{eq-tomonagashift}), provided we set
$p_{\max}=\alpha^{-1}=k_{F}$ and again neglect the ${\cal O}(\omega_{q})$
terms.

Our main result is (\ref{eq-realshift}), the
expressions for the energy shift.
It is interesting to note that (\ref{eq-realshift}) remains regular---without
a pole
or a branch point---at $\Delta=\epsilon_{F}$. When $\Delta_{0,1}=\epsilon_{F}$,
we are in a slightly different regime from the one we have considered, because
some $n_{\rho}=0$, $\ell=\pm1$ states of the trap will be occupied. However,
we can still draw some conclusions from the behavior of the energy shift.
Although the ground state
changes qualitatively at this point, as fermions ``spill over'' out of the
Luttinger liquid into the radially excited state, the phonon energies remain
finite. Even if the decay of a phonon into a $n_{\rho}=0$, $\ell=0$ hole and a
$n_{\rho}=1$, $\ell=0$ fermion is allowed energetically, the phonons appear to
remain metastable against this decay channel.
This represents an interesting
``softening'' of the expected phase transition at this point. However, in
order to understand this situation properly, other interactions than
(\ref{eq-interaction}), including those involving the occupied $\ell=\pm1$
states, must also be considered.

Since the cutoff parameter $\alpha$ is related to the renormalization group
flow for the Luttinger liquid~\cite{ref-chui},
the identification of $k_{F}$ with $\alpha^{-1}$
has interesting implications. In the Luttinger model, $\alpha$ is effectively
a free parameter; its value needs to be specified on the basis of physical
considerations external to the model.
Our results at $T=0$ suggest that $v_{F}k_{F}=2
\epsilon_{F}$ is the maximum physically meaningful bandwidth for the system.
This order of magnitude for the cutoff has a clear physical basis, as discussed
in Section~\ref{sec-bosonized}. Since the Luttinger model (with spin) is known
to be equivalent to several other condensed matter models
\cite{ref-chui,ref-bethe,ref-luther3,ref-luther4,ref-luther5,ref-haldane3} and
field theory models~\cite{ref-heidenreich,ref-coleman},
the specification of a particular value of $\alpha$ may have interesting
implications for the related cutoffs of these analogous systems.

Although we have identified $\alpha^{-1}=k_{F}$ as the correct physical
cutoff at $T=0$, our results do not tell us anything about the renormalization
group flow for nonzero temperatures. Moreover, although (\ref{eq-Tneq0E})
is valid for
finite temperatures, it is only meaningful in an extremely singluar limit; all
the temperature-dependence of the expression has been removed by making
$\Delta$ very large. The temperature-independent expression (\ref{eq-Tneq0E})
is consistent with a nontrivial
$T$-dependence of the energy shift, becuase $\Delta E$ is actually
formally vanishing in this limit, since it is proportional to the ratio
$\left[|g(q)|^{2}\alpha^{-1}q\right]/\Delta$. So the questions of the
finite-temperature energy shift and the nature of the renormalization group
flow for $T>0$ remain open.


We have used several complementary techniques to evaluate the energy shift. The
bosonized method used in Section~\ref{sec-bosonized} is valid in a wider
range of situations; however, we needed the calculations of
Section~\ref{sec-physical} to determine the unknown constant $p_{{\rm max}}$
appearing the bosonized result. We found a frequency shift that is always
${\cal O}[|g(q)|^{2}qk_{F}/\Delta]$, even for nonzero temperatures or in the
presence of strong
Tomonaga-Luttinger-type interactions. When experimenters are able to build 
fermionic traps with sufficiently
high aspect ratios to observe Luttinger liquid behavior and to trap
fermions at sufficiently low temperatures,
formulas such as these should become readily testable.

\section*{Acknowledgments}
The author is grateful to K. Huang for many helpful discussions.

\appendix

\section{Appendix: Harmonic oscillator matrix elements}

\label{sec-appHO}

To obtain the general formula (\ref{eq-HOelements}), we first apply the
Baker-Campbell-Hausdorff formula to the operator $e^{\lambda A^{\dag}-\lambda
^{*}A}$. Since $[-\lambda^{*}A,\lambda A^{\dag}]=-|\lambda|^{2}$ commutes with
both $A$ and $A^{\dag}$, we have
\begin{equation}
\label{eq-BCH}
e^{\lambda A^{\dag}-\lambda^{*}A}=e^{-\lambda^{*}A}e^{\lambda A^{\dag}}
e^{|\lambda|^{2}/2}.
\end{equation}
This reduces the problem to the determination of $\langle n|e^{-\lambda^{*}A}
e^{\lambda A^{\dag}}|m\rangle$.

There are three cases: $m=n$, $m>n$, and $m<n$. The first case is the
simplest. If $m=n$, each factor of $A$ from $e^{-\lambda^{*}A}$ must be paired
with a factor of $A^{\dag}$ from $e^{\lambda A^{\dag}}$; otherwise, the term
does not contribute. So we get
\begin{eqnarray}
\langle m|e^{-\lambda^{*}A}e^{\lambda A^{\dag}}|m\rangle & = & \langle m|1+
\left(-\lambda^{*}A\right)\left(\lambda A^{\dag}\right)+\frac{1}{(2!)^{2}}
\left(-\lambda^{*}A\right)^{2}\left(\lambda A^{\dag}\right)^{2}+\cdots|m\rangle
\nonumber\\
& = & \sum_{i=0}^{\infty}(-1)^{i}|\lambda|^{2i}\frac{1}{(i!)^{2}}\frac{(m+i)!}
{m!} \nonumber\\
\label{eq-l=0}
& \equiv & F\left(m+1;1;-|\lambda|^{2}\right).
\end{eqnarray}

For the case of $m>n$, there must be $l$ more factors of $A^{\dag}$ than
factors of $A$ for a term to contribute. This immediately leads to the series
\begin{eqnarray}
\langle m|e^{-\lambda^{*}A}e^{\lambda A^{\dag}}|n\rangle & = &  \langle m|
\frac{1}{l!}\left(\lambda A^{\dag}\right)^{l}+\frac{1}{(l+1)!}\left(-\lambda
^{*}A\right)\left(\lambda A^{\dag}\right)^{l+1} \nonumber\\
& & +\frac{1}{2!}\frac{1}{(l+2)!}
\left(-\lambda^{*}A\right)^{2}\left(\lambda A^{\dag}\right)^{l+2}+\cdots|n
\rangle \nonumber\\
& = & \lambda^{l}\frac{1}{l!}\sum_{i=0}^{\infty}\frac{1}{i!}\frac{l!}{(l+i)!}
(-1)^{i}|\lambda|^{2i}\frac{(m+i)!}{m!}\sqrt{\frac{m!}{n!}} \nonumber\\
\label{eq-l>0}
& = & \lambda^{l}\frac{1}{l!}\sqrt{\frac{m!}{n!}}F\left(m+1;l+1;-|\lambda|^{2}
\right).
\end{eqnarray}

In the third case, $l$ is negative. The calculation proceeds along essentially
the same lines as for the $l>0$ case. There must be $|l|$ more factors of $A$
than $A^{\dag}$ for a term to be nonzero, so we get
\begin{eqnarray}
\langle m|e^{-\lambda^{*}A}e^{\lambda A^{\dag}}|n\rangle & = &  \langle m|
\frac{1}{|l|!}\left(-\lambda^{*}A\right)^{|l|}+\frac{1}{(|l|+1)!}\left(-\lambda
^{*}A\right)^{|l|+1}\left(\lambda A^{\dag}\right) \nonumber\\
& & +\frac{1}{2!}\frac{1}{(|l|+2)!}
\left(-\lambda^{*}A\right)^{|l|+2}\left(\lambda A^{\dag}\right)^{2}+\cdots|n
\rangle \nonumber\\
& = & \left(-\lambda^{*}\right)^{|l|}\frac{1}{|l|!}\sum_{i=0}^{\infty}\frac{1}
{i!}\frac{|l|!}{(|l|+i)!}(-1)^{i}|\lambda|^{2i}\frac{(n+i)!}{n!}\sqrt{\frac{n!}
{m!}} \nonumber\\
\label{eq-l<0}
& = & \left(-\lambda^{*}\right)^{|l|}\frac{1}{|l|!}\sqrt{\frac{n!}{m!}}F\left(
n+1;|l|+1;-|\lambda|^{2}\right).
\end{eqnarray}
Combining equations (\ref{eq-BCH}), (\ref{eq-l=0}), (\ref{eq-l>0}), and
(\ref{eq-l<0}), we get (\ref{eq-HOelements}).

\section{Appendix: Momentum dependence of $p'_{\max}$}

\label{sec-appcutoff}

We shall now examine how the cutoff $p'_{\max}$ depends upon $q$. We do this
by modifying the prescription (\ref{eq-cutofffind}) for $p'_{\max}$ to
\begin{equation}
\int_{\frac{p_{\min}}{e^{\gamma}}}^{+\infty}\frac{dp_{i}}{p_{i}}\frac{1}{\Delta
+\frac{1}{2m}[(k_{F}-sq-p_{i})^{2}-k_{F}^{2}]+s\omega_{q}+\omega_{p_{i}}}
\approx\frac{1}{\Delta+\frac{q^{2}}{2m}}\log\left(\frac{p'_{\max}}{p_{\min}}
\right).
\end{equation}
To do this, we use the integration formula
\begin{eqnarray}
\label{eq-intformula}
& &\int_{\frac{p_{\min}}{e^{\gamma}}}^{+\infty}\frac{dp_{i}}{ap_{i}+bp_{i}^{2}+
cp_{i}^{3}}=
\frac{1}{a}\log\left(\frac{e^{\gamma}}{p_{{\rm min}}}\sqrt{\frac{a+b
e^{-\gamma}p_{{\rm min}}+
ce^{-2\gamma}p_{{\rm min}}^{2}}{c}}\right) \\
& &+\frac{b}{2a\sqrt{-b^{2}+4ac}}\left[i\log\left(\frac{2ic}{\sqrt{-b^{2}+
4ac}}\right)\!-i\log\left(\frac{-2ic}{\sqrt{-b^{2}+4ac}}\right)\right.
\nonumber\\
& & +i\log\left(\frac
{\sqrt{-b^{2}+4ac}-ib-2ice^{-\gamma}p_{{\rm min}}}{\sqrt{-b^{2}+4ac}}\right)
\left.\!-i\log\left(
\frac{\sqrt{-b^{2}+4ac}+ib+2ice^{-\gamma}p_{{\rm min}}}{\sqrt{-b^{2}+4ac}}
\right)\right]\!\!. \nonumber
\end{eqnarray}
In this case, we have $a=\Delta+\frac{q^{2}}{2m}$, $b=\frac{sq}{m}$, and $c=
\frac{1}{2m}$. We shall define $x$ to be the frequently-appearing ratio
$x\equiv-\frac{b}{\sqrt{-b^{2}+4ac}}=-s\frac{q}{\sqrt{2m\Delta}}$. We shall
find that $p'_{\max}$ depends upon $q$ only through $x$.

To simplify the expression (\ref{eq-intformula}), we first note that since
$p_{{\rm min}}$ is small compared to any other momentum scale, we may
approximate $a+be^{-\gamma}p_{{\rm min}}+ce^{-2\gamma}p_{{\rm min}}^{2}\approx
a$ and $ib+2ice^{-\gamma}p_{{\rm min}}\approx ib$. This reduces the first term
on the right-hand side
to $\frac{1}{a}\log\left(\frac{e^{\gamma}}{p_{{\rm min}}}\sqrt{\frac{a}{c}}
\right)$.

We now turn our attention to the complex terms. Since these terms appear in
conjugate pairs, we need only calculate their real parts.  Each
term has an overall factor of $i$ times a logarithm, so we shall only need the
logarithms' imaginary parts. Because $\frac{2ic}{\sqrt{-b^{2}+4ac}}$
is purely imaginary, we may simplify the second and third terms on the
right-hand side of (\ref{eq-intformula}) to
\begin{eqnarray}
-\frac{x}{2a}\left[i\log\left(\frac{2ic}{\sqrt{-b^{2}+
4ac}}\right)-i\log\left(\frac{-2ic}{\sqrt{-b^{2}+4ac}}\right)\right] & = &
-\frac{x}{2a}\left[i\log i-i\log(-i)\right] \nonumber\\
& = & -\frac{x}{2a}\left[i\left(i\frac{\pi}{2}\right)-i\left(-i\frac{\pi}{2}
\right)\right] \nonumber\\
& = & \frac{\pi x}{2a}.
\end{eqnarray}
We may evaluate the last two terms of (\ref{eq-intformula}) similarly, getting
\begin{eqnarray}
-\frac{x}{2a}\left[\log(1+ix)-i\log(1-ix)\right] & = & -\frac{x}{2a}i\left[
i\tan^{-1}x-i\tan^{-1}(-x)\right] \nonumber\\
& = & \frac{x}{a}\tan^{-1}x.
\end{eqnarray}

So the entire expression becomes
\begin{equation}
\label{eq-cutoffint}
\int_{\frac{p_{\min}}{e^{\gamma}}}^{+\infty}\frac{dp_{i}}{ap_{i}+bp_{i}^{2}+c
p_{i}^{3}}=
\frac{1}{a}\log\left(\frac{e^{\gamma}L\sqrt{2m\Delta+q^{2}}}{2\pi}\right)
+\frac{1}{a}\left[x\left(\tan^{-1}x+\frac{\pi}{2}\right)\right].
\end{equation}
We may absorb the second term on the right-hand side of (\ref{eq-cutoffint})
into the logarithm by exponentiating it. We then identify the cutoff as
\begin{equation}
p'_{{\rm max}}=e^{\gamma}\sqrt{2m\Delta}\sqrt{1+x^{2}}\exp
\left[x\left(\tan^{-1}x+\frac{\pi}{2}\right)\right].
\end{equation}
This expression depends upon $s$. However, only the value of $p'_{\max}$
averaged over $s=\pm1$ is ultimately relevant, so we may modify our
expression for $p'_{\max}$ to become
\begin{equation}
\label{eq-cutoff}
p'_{\max}=e^{\gamma}\sqrt{2m\Delta}\sqrt{1+x^{2}}\exp\left(x\tan^{-1}x\right)
\cosh\left(\frac{\pi x}{2}\right).
\end{equation}
For fixed $\Delta$, this takes on its minimum value in the $q=0$ case
considered previously.

\end{document}